\documentclass[letterpaper]{article} 
\usepackage{aaai2026}  
\usepackage{times}  
\usepackage{helvet}  
\usepackage{courier}  
\usepackage[hyphens]{url}  
\usepackage{graphicx} 
\usepackage{natbib}  
\usepackage{caption} 
\usepackage{algorithm}
\usepackage{algorithmic}

\usepackage{newfloat}
\usepackage{listings}
\DeclareCaptionStyle{ruled}{labelfont=normalfont,labelsep=colon,strut=off} 
\floatstyle{ruled}
\newfloat{listing}{tb}{lst}{}
\floatname{listing}{Listing}
\usepackage{enumitem}
\usepackage{booktabs}

\usepackage{amsmath}
\usepackage{amsthm}
\usepackage{amsfonts}

\newtheorem{theorem}{Theorem}
\newtheorem{remark}{Remark}

\title{DCHO: A Decomposition–Composition Framework for Predicting Higher-Order Brain Connectivity to Enhance Diverse Downstream Applications}

\author{
    Weibin Li\equalcontrib,
    Wendu Li\equalcontrib,
    Quanying Liu\thanks{Corresponding author.},
}
\affiliations{
    Department of Biomedical Engineering, Southern University of Science and Technology, China\\
    liwb2023@mail.sustech.edu.cn, 12432838@mail.sustech.edu.cn, liuqy@sustech.edu.cn
}

\usepackage{bibentry}

\begin{document}

\maketitle

\begin{abstract}
  Higher-order brain connectivity (HOBC), which captures interactions among three or more brain regions, provides richer organizational information than traditional pairwise functional connectivity (FC). Recent studies have begun to infer latent HOBC from noninvasive imaging data, but they mainly focus on static analyses, limiting their applicability in dynamic prediction tasks. To address this gap, we propose DCHO, a unified approach for modeling and forecasting the temporal evolution of \textbf{HO}BC based on a \textbf{D}ecomposition–\textbf{C}omposition framework, which is applicable to both non-predictive tasks (state classification) and predictive tasks (brain dynamics forecasting). DCHO adopts a decomposition–composition strategy that reformulates the prediction task into two manageable subproblems: HOBC inference and latent trajectory prediction. In the inference stage, we propose a dual-view encoder to extract multiscale topological features and a latent combinatorial learner to capture high-level HOBC information. In the forecasting stage, we introduce a latent-space prediction loss to enhance the modeling of temporal trajectories. Extensive experiments on multiple neuroimaging datasets demonstrate that DCHO achieves superior performance in both non-predictive tasks (state classification) and predictive tasks (brain dynamics forecasting), significantly outperforming existing methods. 
\end{abstract}

\begin{links}
    \link{Code}{https://github.com/aaa493/DCHO}
    \link{Data}{https://huggingface.co/datasets/aaa493/DCHO}
\end{links}

\section{Introduction}
Functional connectivity (FC) \cite{yeo2011organization, hutchison2013dynamic}, the most widely used framework, models pairwise statistical dependencies between brain regions based on fMRI data. FC is limited to pairwise interactions and may overlook higher-level coordination. Recent studies highlight the importance of higher-order brain connectivity (HOBC) that involves three or more regions, supported by evidence across scales \cite{BATTISTON20201, battiston2022higher, bianconi2021higher, jun2017fully, steinmetz2021neuropixels, paulk2022large, chelaru2021high, tadic2022multiscale}. In human neuroimaging, direct observation of HOBC remains challenging due to the limitations of noninvasive techniques. A recent study \cite{santoro2024higher, santoro2023higher} was the first to infer latent HOBC from fMRI signals and applied them to downstream tasks such as classification. However, they focus on analyzing HOBC within static or predefined temporal windows, without modeling their temporal evolution. This limits the potential of HOBC in predictive applications, such as forecasting brain dynamics. Addressing this gap requires models that can anticipate the dynamic trajectory of the HOBC over time.

\begin{figure}[t]
\begin{center}
\centerline{\includegraphics[width=1.0\columnwidth]{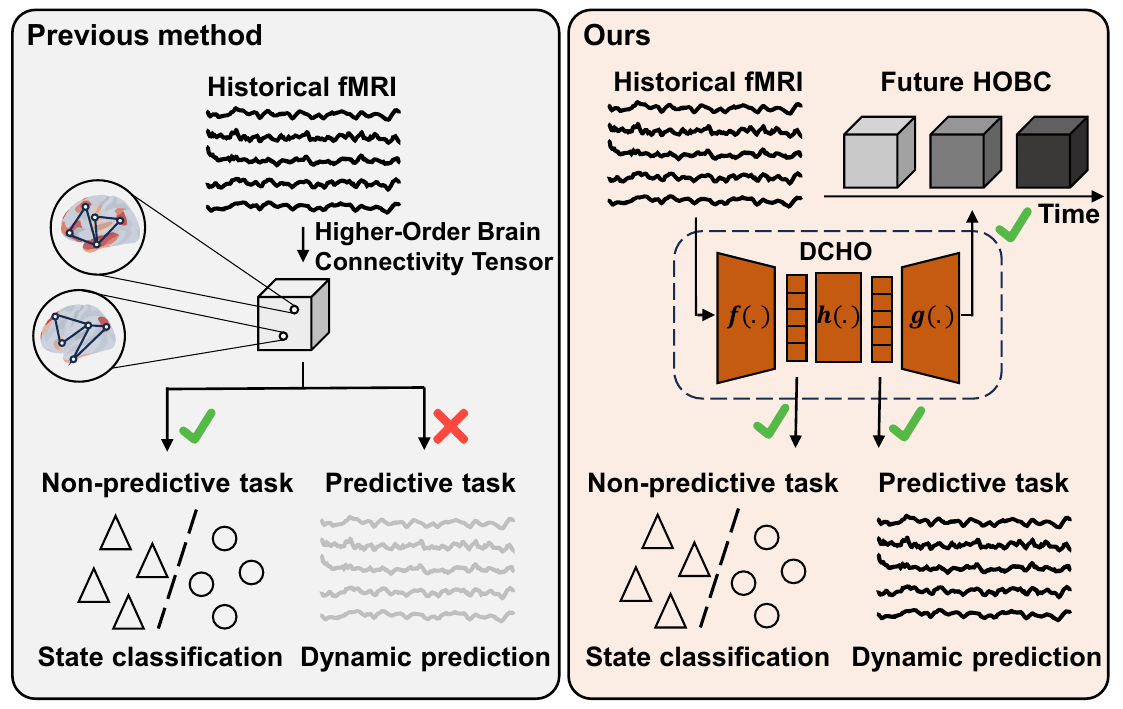}}
\caption{\textbf{Motivation.} (Left): Previous methods analyze HOBC in static windows, overlooking their temporal evolution and limiting predictive applications. (Right): DCHO overcomes this limitation by forecasting the dynamic trajectories of HOBC.}
\label{fig1}
\end{center}
\end{figure}

Predicting the evolution of the HOBC in complex brain systems is challenging. First, the combinatorial explosion in the number of possible higher-order interactions poses a major obstacle to direct modeling \cite{bassett2017network, breakspear2017dynamic, ROEBROECK2011296, HERZOG2022105918}. The curse of dimensionality poses a major challenge to inferring the HOBC. Second, brain dynamics are inherently nonlinear, nonstationary, and stochastic, governed by latent neural processes that evolve across multiple spatial and temporal scales. Capturing the temporal evolution of HOBC thus requires not only modeling individual node trajectories, but also learning coordinated multi-region dependencies that reflect complex integration, segregation, and recurrent feedback mechanisms. These interactions are not additive but shaped by nonlinear dynamical couplings that defy simple compositional rules \cite{Topologyshapes, battiston2021physics, zhang2023higher, alvarez2021evolutionary}. 

To address the above limitations, we propose DCHO, a model designed to capture and predict the temporal evolution of higher-order brain connectivity, while leveraging its high-level representations to support both non-predictive and predictive tasks. DCHO introduces an innovative decomposition–composition framework that reformulates the complex prediction task into two more tractable subtasks: HOBC tensor inference and latent trajectory prediction. In the inference stage, DCHO employs a dual-view encoder to capture multi-scale topological features and incorporates a latent composition learner to further encode higher-order interaction patterns, effectively mitigating the curse of dimensionality and enhancing representational capacity. In the forecasting stage, DCHO introduces a latent-space prediction loss to model the dynamic evolution of HOBC in an abstract information space. With the pre-trained encoder and predictor, DCHO is broadly applicable to both non-predictive tasks (e.g., state classification) and predictive tasks (e.g., brain dynamics forecasting), and consistently outperforms existing state-of-the-art methods across multiple datasets (Figure \ref{fig1}). Our main contributions are summarized as follows:
\begin{itemize}
    \item \textbf{Decomposition–Composition Framework:} DCHO decomposes prediction into HOBC inference and latent forecasting, supported by theoretical analysis and enhanced by a latent-space prediction loss for abstract trajectory modeling.
    \item \textbf{Dual-view Encoder and Latent Combinatorial Learner:} We propose the dual-view encoder and the latent learner to capture multiscale topologies and high-level HOBC information.
    \item \textbf{Effective Across Predictive and Non-predictive Applications:} DCHO leverages pretrained encoder and predictor for both non-predictive (state classification) and predictive tasks (brain dynamics forecasting), outperforming state-of-the-art methods on multiple datasets.
\end{itemize}

\section{Related Work}
\textbf{Higher-order Brain Connectivity: }Although FC \cite{yeo2011organization, hutchison2013dynamic} has been the standard for brain network modeling, it fails to capture higher-order interactions among multiple brain regions. HOBC describes complex coordination among three or more regions and reveals richer organizational patterns. However, due to the limitations of non-invasive neuroimaging techniques, directly observing HOBC in humans remains challenging. Recent work \cite{santoro2024higher, santoro2023higher} has attempted to infer HOBC from fMRI and apply it to static classification tasks. However, the study overlooked its temporal dynamics, limiting their use in predictive settings. To bridge this gap, we propose the DCHO, which models and forecasts the temporal evolution of HOBC from neural signals, enabling both non-predictive tasks (task state classification) and more demanding predictive tasks (brain dynamics forecasting).\\
\textbf{Brain Dynamics Forecasting: }
In recent years, various architectures have been proposed for modeling neural dynamics. RNN-based methods (LtrRNN \cite{pellegrino2023low} and LSTM \cite{graves2012long}) are effective in capturing short-term temporal dependencies but are limited in expressing long-term dynamics. Transformer-based models (NetFormer \cite{lu2025netformer}, STNDT \cite{le2022stndt} and Transformer \cite{vaswani2017attention}) leverage self-attention mechanisms to better model global temporal dependencies, yet often overlook the underlying brain connectivity structure. GNN-based methods (AMAG \cite{li2023amag}) are naturally suited for encoding spatial topological structures between brain regions and have been applied to model static or pairwise functional connectivity, but typically neglect global topological information. In contrast, the proposed DCHO framework integrates a dual-graph encoder and a latent learner to capture high-level information representations of HOBC and explicitly models the underlying dynamic evolution in the latent space, substantially enhancing modeling capacity and robustness in long-term prediction.

\begin{figure*}[t]
\centering
\includegraphics[width=1.0\textwidth]{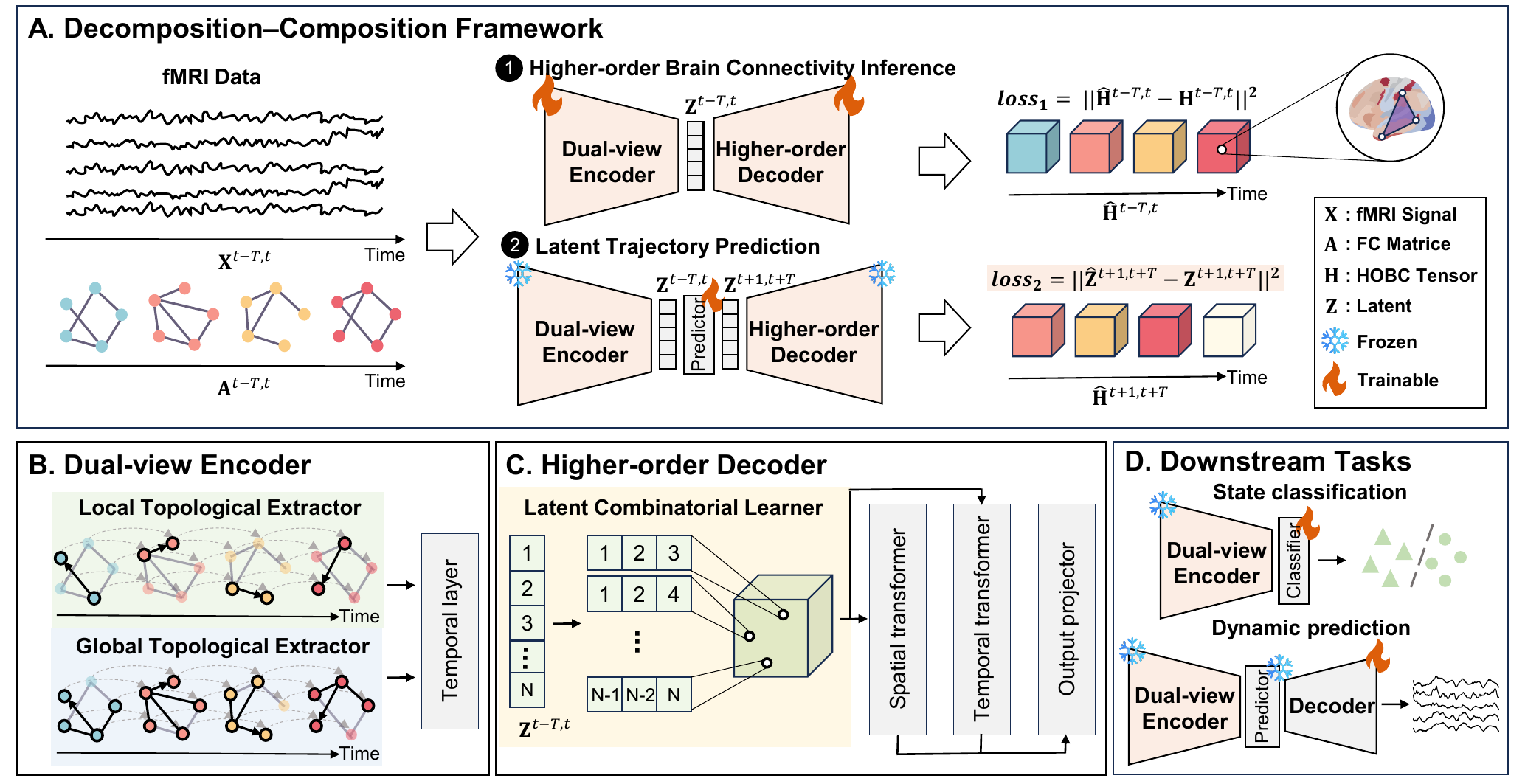}
\caption{\textbf{Overview of DCHO} (A) \textbf{Decomposition–Composition Framework:} DCHO framework decomposes prediction into HOBC inference and latent trajectory prediction, using a \textbf{latent-space prediction loss} to model high-level temporal dynamics. (B) \textbf{Dual-view Encoder: }DCHO applies two parallel GNN branches that extract local and global topological features. (C) Higher-order Decoder: DCHO proposes a \textbf{latent combinatorial learner} to capture high-level HOBC information. (D) DCHO leverages the pretrained encoder and predictor to support both non-predictive and predictive tasks.}
\label{fig2}
\end{figure*}

\section{Preliminaries}
\subsection{The Definition of Higher-order Brain Connectivity}
We first introduce the definition of HOBC \cite{santoro2024higher, santoro2023higher}. Let $\mathbf{x}_i \in \mathbb{R}^T = [x_i^1, x_i^2, \ldots, x_i^T]$ denotes the original time series of region $i$. We first $z$-score each region’s time series:
\begin{equation}
\mathbf{\tilde{x}}_i = \frac{\mathbf{x}_i - \mu[\mathbf{x}_i]}{\sigma[\mathbf{x}_i]},
\end{equation}
where $\mu[\cdot]$ and $\sigma[\cdot]$ denote the time-averaged mean and standard deviation, respectively. For a $(k+1)$-node simplex, we define the $k$-order $z$-scored co-fluctuation signal at time $t$ as:
\begin{equation}
\xi_{0 \ldots k}^t = 
\frac{
\prod_{p=0}^{k} \tilde{x}_{p}^t - \mu \left[ \prod_{p=0}^{k} \mathbf{\tilde{x}}_{p} \right]
}{
\sigma \left[ \prod_{p=0}^{k} \mathbf{\tilde{x}}_{p} \right]
}.
\end{equation}

To distinguish concordant from discordant interactions within a $k$-order product, we assign positive values to fully concordant signs, whereas discordant combinations are mapped to negative values. Namely, 
\begin{equation}
\mathrm{sign}\left[ \xi_{0 \ldots k}^t \right] := (-1)^{\mathrm{sgn} \left( (k+1) - \left| \sum_{0}^{k} \mathrm{sgn}[\tilde{x}_i^t] \right| \right)},
\end{equation}
where $\mathrm{sgn}[\cdot]$ is the signum function of a real number. The final weighted co-fluctuation signal is defined as:
\begin{equation}
w_{0 \ldots k}^t = 
\text{sign} \left[ \xi_{0 \ldots k}^t \right] 
\cdot \left| \xi_{0 \ldots k}^t \right|.
\end{equation}
If all $k$-order products are computed, this yields a total of $\binom{N}{k+1}$ distinct co-fluctuation time series for each order $k$, where $N$ denotes the number of brain regions.

At each time point $t$, we organize the resulting $k$-order co-fluctuations into a weighted simplicial complex $\mathcal{K}^t$. For simplicity, in this work, we only consider co-fluctuations of dimension up to $k = 2$, so that triangles represent the higher-order connectivity in the weighted simplicial complex $\mathcal{K}$. We represent the higher-order connectivity \( \mathcal{K}^t \) as a third-order tensor \( H^t \in \mathbb{R}^{N \times N \times N} \), where each element \( h_{ijk}^{t} \) denotes the weighted co-fluctuation signal \( w_{ijk}^t \) among regions \( i, j, k \) at time \( t \).

\section{Method}
\subsection{Decomposition–Composition Framework}
Let $\mathbf{X}^{t-T, t} = [X^{t-T}; \ldots; X^{t}] \in \mathbb{R}^{T \times N}$ be the original fMRI signals and $\mathbf{A}^{t-T, t} = [A^{t-T}; \ldots; A^{t}] \in \mathbb{R}^{T \times N \times N}$ be the corresponding functional connectivity matrices. Our goal is to predict the future value of HOBC $\hat{\mathbf{H}}^{t+1, t+T} = [\hat{H}^{t+1}; \ldots; \hat{H}^{t+T}] \in \mathbb{R}^{T \times N \times N \times N}$ based on $\mathbf{A}^{t-T, t}$ and $\mathbf{X}^{t-T, t}$. In this study, we present a decomposition–composition framework that breaks the overall objective into two tractable subtasks: (1) higher-order brain connectivity inference and (2) latent trajectory prediction. The overview of DCHO is summarized in Figure \ref{fig2} and Algorithm \ref{alg1}. Algorithm 1 is provided in the Appendix \ref{Algorithm}.

The first subtask aims to infer a sequence of the HOBC tensors $\hat{\mathbf{H}}^{t-T, t}$ based on $\mathbf{X}^{t-T, t}$ and $\mathbf{A}^{t-T, t}$. A dual-view encoder first transforms these inputs into latent spatio-temporal representations $\mathbf{Z}^{t-T, t}$ by applying two parallel GNN branches that extract local and global topological features. These latent representations are then fed into a higher-order decoder to infer the HOBC tensors. The decoder consists of a latent combinatorial learner and a dual-stream Transformer module, where the learner further encourages the latent representation $\mathbf{Z}^{t-T, t}$ to capture higher-order interaction patterns. The loss of the first subtask is:
\begin{equation}
loss_{1} = ||\hat{\mathbf{H}}^{t-T, t} - \mathbf{H}^{t-T, t}||^2.
\end{equation}

The second subtask focuses on predicting future latent dynamics. We apply a LSTM-based predictor to model temporal dependencies in the latent space and predict future embeddings $\hat{\mathbf{Z}}^{t+1,t+T}$ from $\mathbf{Z}^{t-T,t}$. This latent trajectory prediction forms the basis for estimating future HOBC tensors. The loss of the second subtask is:
\begin{equation}
loss_{2} = ||\hat{\mathbf{Z}}^{t+1, t+T} - \mathbf{Z}^{t+1, t+T}||^2.
\end{equation}

The training process is conducted in two stages: we first train the higher-order inference subtask, and subsequently freeze both the encoder and decoder to optimize the latent trajectory prediction subtask. The advantages of the proposed framework are theoretically justified and analyzed in the following. 

\begin{theorem}
Let \( f_{\mathrm{enc}}: \mathbb{R}^{T \times N \times N} \to \mathbb{R}^{T \times N \times F}\), \( f_{\mathrm{dyn}}: \mathbb{R}^{T \times N \times F} \to \mathbb{R}^{T \times N \times F}\), and \( f_{\mathrm{dec}}: \mathbb{R}^{T \times N \times F} \to \mathbb{R}^{T \times N \times N \times N} \) be measurable mappings that are Lipschitz continuous, with Lipschitz constants \( L_{\mathrm{enc}}, L_{\mathrm{dyn}}, L_{\mathrm{dec}} \), respectively. The prediction of a trained model is defined as:
\begin{equation}
\hat{\mathbf{H}}^{t+1, t+T}=f_{\mathrm{dec}}\bigl(
    f_{\mathrm{dyn}}\bigl(
       f_{\mathrm{enc}}(\mathbf{X}^{t-T, t}, \mathbf{A}^{t-T, t},)
    \bigr)
\bigr).
\end{equation} 
The inference error is defined as: 
\begin{equation}
\epsilon_{\mathrm{inf}}=\bigl\|f_{\mathrm{dec}}\bigl(f_{\mathrm{enc}}(\mathbf{X}^{t+1, t+T}, \mathbf{A}^{t+1, t+T})\bigr) - \mathbf{H}^{t+1, t+T}\bigr\|.
\end{equation}
Let $\mathbf{Z}^{t-T,t} =  f_{\mathrm{enc}}(\mathbf{X}^{t-T, t}, \mathbf{A}^{t-T, t})$. The latent prediction error is defined as: 
\begin{equation}
\epsilon_{\mathrm{dyn}}
=
\bigl\|
    f_{\mathrm{dyn}}\bigl(
        \mathbf{Z}^{t-T,t} 
    \bigr)
    -
    \mathbf{Z}^{t+1, t+T} 
\bigr\|.
\end{equation}
Then the prediction error satisfies:
\begin{equation}
\bigl\|
    \hat{\mathbf{H}}^{t+1, t+T}
    -
    \mathbf{H}^{t+1, t+T}
\bigr\| \le \epsilon_{\mathrm{inf}} + L_{\mathrm{dec}}\cdot \epsilon_{\mathrm{dyn}}.
\end{equation}
\end{theorem}

The proof of Theorem 1 is provided in the Appendix \ref{The Proof of Theorem 1}.

\begin{remark}
By decoupling the sources of error, each component can be independently analyzed and optimized, making the total prediction error more controllable and theoretically better bounded than in end-to-end models with entangled errors. Benefiting from the dual-view encoder, latent combinatorial learner, and latent-space prediction loss, DCHO effectively reduces both the inference error \(\epsilon_{\mathrm{inf}}\) and the latent dynamics error \(\epsilon_{\mathrm{dyn}}\), enabling fine-grained modeling of the two subtasks. As shown in Table \ref{ablation}, ablation results further validate the effectiveness of the decomposition–composition framework.
\end{remark}

\subsection{The Dual-view Encoder}

The dual-view encoder consists of two parallel GNN branches---a local and a global topological extractor---that process spatiotemporal inputs $\mathbf{X}^{t-T,t}$ and $\mathbf{A}^{t-T,t}$ into latent representations $\mathbf{Z}^{t-T,t}$ enriched with structural and temporal information. Details are as follows:

(i) \textbf{The local topological extractor} adaptively encodes pairwise interactions using a GNN-based mechanism \cite{kipf2017semisupervisedclassificationgraphconvolutional, xu2019powerful}. First, interaction scores are computed between the central node and its neighbors, and these scores are used to weight the neighbors' embeddings from the previous layer. 
Mathematically, for the embedding of the $i_t$ node $v_i^t$ at the $s$-th layer: $e_i^{t,(s)}$, the interaction scores between the $i_t$ node and its neighbor $j_t$ are derived from the adjacency matrix and the node features in the temporal graph: 
\begin{align}\label{eq3.1}
    {sc}^{(s)}(v_i^t, v_j^t) = A(i_t, j_t) \cos\left( W_{\text{q}} e_i^{t,(s)}, W_{\text{k}} e_j^{t,(s)} \right),
\end{align}
where the cosine similarity: \( \cos(\cdot, \cdot) \) quantifies the similarity between two vectors. \( W_{\text{q}} \) and \( W_{\text{k}} \) are transformation matrices employed for similarity calculations. Then, the node representation at the \((s+1)\)-th layer is updated as follows:
\begin{align}\label{eq3.2}
    e_i^{t, (s+1)} = e_i^{t, (s)} + \sigma \left( \sum_{v_j^{t} \in N_{v_i^t}} {sc}^{(s)}(v_i^t, v_j^t) W_{\text{v}} e_j^{t, (s)} \right),
\end{align}
where \( \sigma(\cdot) \) is a non-linear activation function, \( W_{\text{v}} \) is a learnable matrix, and \( N_{v_i^t} \) denote as the neighbors of \( v_i^t \). After \( S \) layers, the final node representation: \( e_i^{t} = e_i^{t, (S)} \) is adaptively determined.

(ii) \textbf{The global topological extractor} is designed based on spectral graph convolution, which serves as a complementary approach to uncover HOBC information hidden in the spectral domain. Specifically, first, we introduce the necessary mathematical definitions. The Chebyshev polynomials are defined as \(T_0(x) = 1\), \(T_1(x) = x\), and \(T_m(x) = 2xT_{m-1}(x) - T_{m-2}(x)\). The normalized graph Laplacian $L$ is defined as \(L = I - \widetilde{D}^{-\frac{1}{2}} A \widetilde{D}^{-\frac{1}{2}}\), where $D$ is the degree matrix and $I$ is the identity matrix. Define the matrix \(R^{(0)} = X\), which stacks the features of all nodes. Then, based on the second-order Chebyshev graph convolution \cite{defferrard2016convolutional}, the node representation matrix \(R^{(s)}\) at the $s$-th layer is calculated as:
\begin{align}\label{eq3.3}
    R^{(s)} = \sum_{m=0}^2 T_m(\widetilde{L})R^{(s-1)}W_m^{(s)}
\end{align}
where \(\widetilde{L} = \frac{2L}{\lambda_{\text{max}}} - I\), \(\lambda_{\text{max}}\) denotes the maximum eigenvalue of \(L\), and \(W_m^{(s)} \in \mathbb{R}^{d \times d}\) is a learnable weight matrix at the \(s\)-th layer. By stacking $S$ layers, we derive each representation vector \(r_i^t\) from \(R^{(S)} \in \mathbb{R}^{NT \times d}\), capturing higher order non-local information from a spectral perspective.

(iii) The above two extractors have captured the representations of pairwise interactions and higher-order interactions, respectively. We then employ an attention mechanism to aggregate these temporal representations into a latent representation $\mathbf{Z}_i^t$ with advanced spatio-temporal features. Specifically, first, we merge the two representations from the last layer with the temporal embeddings $TE(t)$, and apply a MLP \(\delta(\cdot)\) to calculate the representation \(c_i^t\) for each node \(v_i^t\). Mathematically:
\begin{align}\label{eq3.4}
c_i^t &= \delta([e_i^t, r_i^t] + TE(t)).  
\end{align}
We then refine the node representation through an attention-based transformation that integrates contextual dependencies at each timestamp. Specifically, an attention refinement operator is applied to produce a context-aware intermediate state, which is further processed by a position-wise feed-forward network to obtain the final latent representation. Mathematically:
\begin{align}\label{eq3.7}
\zeta_i^{t} &= \psi(c_i^{t}+f_{att}(c_i^{t})),\\\label{eq3.8}
\mathbf{Z}_i^t &= \psi(\zeta_i^{t}+f_{ffn}(\zeta_i^{t})).
\end{align}
where $f_{\mathrm{att}}(\cdot)$ denotes an attention refinement operator implemented with multi-head self-attention, $f_{\mathrm{ffn}}(\cdot)$ is a position-wise feed-forward network, and $\psi(\cdot)$ denotes a normalization operator (Layer Normalization ).

\subsection{Higher-order Decoder}

The higher-order decoder infers a sequence of HOBC tensors $\hat{\mathbf{H}}^{t-T, t}$ from the latent representation $\hat{\mathbf{Z}}^{t-T, t}$, and is mainly composed of a latent combinatorial learner and a dual-stream transformer module. The details are presented below:

(i) \textbf{The latent combinatorial learner:} For each region embedding $Z^{t-T, t}_{i}, Z^{t-T, t}_{j}, Z^{t-T, t}_{k} \in \mathbb{R}^{T \times F}$ from $\mathbf{Z}^{t-T, t}$, we apply three linear projections:
\begin{equation}
    h_i = W_i Z^{t-T, t}_{i}, \quad h_j = W_j Z^{t-T, t}_{j}, \quad h_k = W_k Z^{t-T, t}_{k},
\end{equation}
where $W_i, W_j, W_k \in \mathbb{R}^{F \times F}$ are learnable linear projections. 
These are concatenated and projected to obtain unified triplet tokens:
\begin{equation}
    o^{t-T, t}_{ijk} = \text{Linear} \left( [h_i, h_j, h_k] \right) \in \mathbb{R}^{T, 3 \times F}.
\end{equation}

All $M$ triplet-level representations $o_{ijk}^{t-T,t} \in \mathbb{R}^{T \times 3F}$ are stacked to form the unified tensor $o_{M}^{t-T,t} \in \mathbb{R}^{T \times M \times 3F}$, where $M=\binom{N}{3} = \frac{N(N - 1)(N - 2)}{6}$ denotes the total number of triplets. This tensor is then passed through a linear layer to obtain the final combinatorial representation $\tilde{o}^{t-T, t}_M \in \mathbb{R}^{T, M, F}$.

(ii) \textbf{Spatial and Temporal stream:} We add the positional encoding $\text{PosEnc}(M)$ for each time step $t$, and the temporal encoding $\text{TimeEnc}(T)$ for each triplet $(i, j, k)$:

\begin{align}
    \tilde{o}_{\text{s}} &= \text{Transformer}_{\text{s}} \left( \tilde{o}^{t-T, t}_M + \text{PosEnc}(M) \right), \\
    \tilde{o}_{\text{t}} &= \text{Transformer}_{\text{t}} \left( \tilde{o}^{t-T, t}_M + \text{TimeEnc}(T) \right).
\end{align}

The spatial and temporal streams are then concatenated and passed through an output layer to obtain the final inference.

\begin{equation}
    \hat{\mathbf{H}}^{t-T, t} = \text{MLP} \left( \left[ \tilde{o}_{\text{s}}, \tilde{o}_{\text{s}} \right] \right) \in \mathbb{R}^{T \times N \times N \times N}.
\end{equation}

This dual-stream architecture jointly captures higher-order spatial topologies and their dynamic evolution.

\subsection{LSTM-based predictor}

We employ a multi-layer LSTM network to model the temporal evolution of latent representations. Given the latent states $\mathbf{Z}^{t-T,t}$ from the encoder, the predictor forecasts future trajectories as:
\begin{equation}
\hat{\mathbf{Z}}^{t+1,t+T} = \text{LayerNorm}(\text{Linear}(\text{LSTM}(\mathbf{Z}^{t-T,t}))).
\end{equation}
This architecture captures long-range dependencies in HOBC dynamics, providing informative latent features for future HOBC prediction.

\begin{table*}[h]
\centering
\label{tab:baseline_results}
\resizebox{\textwidth}{!}{%
\begin{tabular}{lcccccccccc}
\toprule
Metrics & \multicolumn{10}{c}{MAE} \\
\cmidrule(lr){1-11}
Dataset & Emotion & Gambling & Language & Motor & Relational & Social & WM & Rest & Lorenz & HR \\
\midrule
MLP & 0.3660 \scriptsize $\pm$ 0.0005 & 0.3008 \scriptsize $\pm$ 0.0009 & 0.3311 \scriptsize $\pm$ 0.0008 & 0.2907 \scriptsize $\pm$ 0.0008 & 0.4939 \scriptsize $\pm$ 0.0052 & 0.3492 \scriptsize $\pm$ 0.0001 & 0.3594 \scriptsize $\pm$ 0.0062 & 0.5875 \scriptsize $\pm$ 0.0003 & 0.1155 \scriptsize $\pm$ 0.0017 & 0.0833 \scriptsize $\pm$ 0.0028 \\
Trans. & 0.3618 \scriptsize $\pm$ 0.0020 & 0.2605 \scriptsize $\pm$ 0.0018 & 0.2532 \scriptsize $\pm$ 0.0010 & 0.2185 \scriptsize $\pm$ 0.0023 & \underline{0.3783} \scriptsize $\pm$ 0.0069 & 0.2844 \scriptsize $\pm$ 0.0005 & 0.2751 \scriptsize $\pm$ 0.0049 & 0.4959 \scriptsize $\pm$ 0.0019 & \underline{0.1250} \scriptsize $\pm$ 0.0024 & 0.0559 \scriptsize $\pm$ 0.0019 \\
LSTM & \underline{0.2887} \scriptsize $\pm$ 0.0004 & \underline{0.2216} \scriptsize $\pm$ 0.0009 & \underline{0.2352} \scriptsize $\pm$ 0.0005 & \underline{0.1971} \scriptsize $\pm$ 0.0009 & 0.4029 \scriptsize $\pm$ 0.0114 & \underline{0.2604} \scriptsize $\pm$ 0.0001 & \underline{0.2488} \scriptsize $\pm$ 0.0033 & \underline{0.3865} \scriptsize $\pm$ 0.0002 & 0.2608 \scriptsize $\pm$ 0.0085 & \underline{0.0416} \scriptsize $\pm$ 0.0009 \\
DCHO & \textbf{0.1744} \scriptsize $\pm$ 0.0004 & \textbf{0.0720} \scriptsize $\pm$ 0.0003 & \textbf{0.1156} \scriptsize $\pm$ 0.0005 & \textbf{0.0778} \scriptsize $\pm$ 0.0004 & \textbf{0.2674} \scriptsize $\pm$ 0.0053 & \textbf{0.1102} \scriptsize $\pm$ 0.0002 & \textbf{0.1391} \scriptsize $\pm$ 0.0011 & \textbf{0.1704} \scriptsize $\pm$ 0.0004 & \textbf{0.0780} \scriptsize $\pm$ 0.0001 & \textbf{0.0249} \scriptsize $\pm$ 0.0001 \\
\midrule
Metrics & \multicolumn{10}{c}{RMSE} \\
\cmidrule(lr){1-11}
Dataset & Emotion & Gambling & Language & Motor & Relational & Social & WM & Rest & Lorenz & HR \\
\midrule
MLP & 0.5911 \scriptsize $\pm$ 0.0022 & 0.4790 \scriptsize $\pm$ 0.0015 & 0.5402 \scriptsize $\pm$ 0.0014 & 0.4589 \scriptsize $\pm$ 0.0024 & 0.8385 \scriptsize $\pm$ 0.0070 & 0.5649 \scriptsize $\pm$ 0.0020 & 0.6057 \scriptsize $\pm$ 0.0049 & 1.1158 \scriptsize $\pm$ 0.0067 & 0.2017 \scriptsize $\pm$ 0.0070 & 0.1772 \scriptsize $\pm$ 0.0055 \\
Trans. & 0.6033 \scriptsize $\pm$ 0.0062 & 0.4466 \scriptsize $\pm$ 0.0020 & 0.4117 \scriptsize $\pm$ 0.0008 & 0.3446 \scriptsize $\pm$ 0.0041 & \underline{0.4971} \scriptsize $\pm$ 0.0177 & 0.4712 \scriptsize $\pm$ 0.0010 & 0.4854 \scriptsize $\pm$ 0.0074 & 0.9793 \scriptsize $\pm$ 0.0030 & \underline{0.1923} \scriptsize $\pm$ 0.0037 & 0.1083 \scriptsize $\pm$ 0.0070 \\
LSTM & \underline{0.4709} \scriptsize $\pm$ 0.0008 & \underline{0.3858} \scriptsize $\pm$ 0.0016 & \underline{0.3874} \scriptsize $\pm$ 0.0010 & \underline{0.3170} \scriptsize $\pm$ 0.0012 & 0.6936 \scriptsize $\pm$ 0.0228 & \underline{0.4314} \scriptsize $\pm$ 0.0007 & \underline{0.4451} \scriptsize $\pm$ 0.0072 & \underline{0.5240} \scriptsize $\pm$ 0.0030 & 0.3779 \scriptsize $\pm$ 0.0269 & \underline{0.0904} \scriptsize $\pm$ 0.0031 \\
DCHO & \textbf{0.2672} \scriptsize $\pm$ 0.0011 & \textbf{0.1161} \scriptsize $\pm$ 0.0009 & \textbf{0.1852} \scriptsize $\pm$ 0.0009 & \textbf{0.1239} \scriptsize $\pm$ 0.0009 & \textbf{0.3298} \scriptsize $\pm$ 0.0267 & \textbf{0.1746} \scriptsize $\pm$ 0.0005 & \textbf{0.2289} \scriptsize $\pm$ 0.0022 & \textbf{0.2738} \scriptsize $\pm$ 0.0042 & \textbf{0.1388} \scriptsize $\pm$ 0.0288 & \textbf{0.0412} \scriptsize $\pm$ 0.0186 \\
\bottomrule
\end{tabular}
}
\caption{Prediction performance comparison (MAE and RMSE) of DCHO and baseline models on multiple datasets for HOBC forecasting with a 10-step horizon.}
\label{table1}
\end{table*}

\begin{table*}[h]
\centering
\label{tab:baseline_results}
\resizebox{\textwidth}{!}{%
\begin{tabular}{lcccccccccc}
\toprule
Metrics & \multicolumn{10}{c}{MAE} \\
\cmidrule(lr){1-11}
Dataset & Emotion & Gambling & Language & Motor & Relational & Social & WM & Rest & Lorenz & HR \\
\midrule
MLP & 0.7649 \scriptsize $\pm$ 0.0008 & 0.7554 \scriptsize $\pm$ 0.0024 & 0.7101 \scriptsize $\pm$ 0.0019 & 0.7469 \scriptsize $\pm$ 0.0024 & 0.6550 \scriptsize $\pm$ 0.0068 & 0.7448 \scriptsize $\pm$ 0.0001 & 0.7355 \scriptsize $\pm$ 0.0090 & 0.7834 \scriptsize $\pm$ 0.0001 & 0.1381 \scriptsize $\pm$ 0.0025 & 0.4927 \scriptsize $\pm$ 0.0060 \\
Trans. & 0.2651 \scriptsize $\pm$ 0.0018 & 0.3176 \scriptsize $\pm$ 0.0022 & 0.2781 \scriptsize $\pm$ 0.0009 & 0.2525 \scriptsize $\pm$ 0.0016 & 0.7803 \scriptsize $\pm$ 0.0110 & 0.2780 \scriptsize $\pm$ 0.0019 & 0.3433 \scriptsize $\pm$ 0.0019 & 0.8444 \scriptsize $\pm$ 0.0018 & 0.1017 \scriptsize $\pm$ 0.0021 & 0.6998 \scriptsize $\pm$ 0.0096 \\
LSTM & 0.2361 \scriptsize $\pm$ 0.0001 & 0.2831 \scriptsize $\pm$ 0.0002 & \underline{0.2625} \scriptsize $\pm$ 0.0003 & \underline{0.2347} \scriptsize $\pm$ 0.0003 & 0.9548 \scriptsize $\pm$ 0.0162 & 0.2461 \scriptsize $\pm$ 0.0001 & \underline{0.2761} \scriptsize $\pm$ 0.0009 & 0.9874 \scriptsize $\pm$ 0.0002 & \underline{0.1033} \scriptsize $\pm$ 0.0008 & 0.3726 \scriptsize $\pm$ 0.0484 \\
STNDT & 0.6624 \scriptsize $\pm$ 0.0052 & 0.7134 \scriptsize $\pm$ 0.0038 & 0.7802 \scriptsize $\pm$ 0.0033 & 0.6964 \scriptsize $\pm$ 0.0023 & 0.7211 \scriptsize $\pm$ 0.0057 & 0.7191 \scriptsize $\pm$ 0.0001 & 0.7119 \scriptsize $\pm$ 0.0010 & 0.7637 \scriptsize $\pm$ 0.0005 & 0.3146 \scriptsize $\pm$ 0.0105 & 0.2961 \scriptsize $\pm$ 0.0454 \\
AMAG & \underline{0.2103} \scriptsize $\pm$ 0.0020 & \underline{0.1973} \scriptsize $\pm$ 0.0019 & 0.3571 \scriptsize $\pm$ 0.0012 & 0.4734 \scriptsize $\pm$ 0.0015 & \underline{0.4820} \scriptsize $\pm$ 0.0062 & \underline{0.1524} \scriptsize $\pm$ 0.0009 & 0.4790 \scriptsize $\pm$ 0.0023 & \underline{0.5558} \scriptsize $\pm$ 0.0009 & 0.5547 \scriptsize $\pm$ 0.0107 & \underline{0.1670} \scriptsize $\pm$ 0.0785 \\
LtrRNN & 0.6665 \scriptsize $\pm$ 0.0077 & 0.7559 \scriptsize $\pm$ 0.0026 & 0.8094 \scriptsize $\pm$ 0.0033 & 0.7811 \scriptsize $\pm$ 0.0036 & 0.7423 \scriptsize $\pm$ 0.0032 & 0.7772 \scriptsize $\pm$ 0.0001 & 0.7870 \scriptsize $\pm$ 0.0009 & 0.7926 \scriptsize $\pm$ 0.0007 & 0.7222 \scriptsize $\pm$ 0.0068 & 0.8560 \scriptsize $\pm$ 0.0588 \\
NetFormer & 0.5873 \scriptsize $\pm$ 0.0026 & 0.6776 \scriptsize $\pm$ 0.0030 & 0.7121 \scriptsize $\pm$ 0.0034 & 0.6367 \scriptsize $\pm$ 0.0016 & 0.6615 \scriptsize $\pm$ 0.0027 & 0.6437 \scriptsize $\pm$ 0.0001 & 0.6343 \scriptsize $\pm$ 0.0009 & 0.7014 \scriptsize $\pm$ 0.0004 & 0.2596 \scriptsize $\pm$ 0.0090 & 0.5355 \scriptsize $\pm$ 0.0860 \\
DCHO & \textbf{0.1582} \scriptsize $\pm$ 0.0011 & \textbf{0.0750} \scriptsize $\pm$ 0.0011 & \textbf{0.1382} \scriptsize $\pm$ 0.0006 & \textbf{0.0936} \scriptsize $\pm$ 0.0008 & \textbf{0.2802} \scriptsize $\pm$ 0.0072 & \textbf{0.0701} \scriptsize $\pm$ 0.0006 & \textbf{0.1801} \scriptsize $\pm$ 0.0009 & \textbf{0.2515} \scriptsize $\pm$ 0.0007 & \textbf{0.0613} \scriptsize $\pm$ 0.0012 & \textbf{0.0545} \scriptsize $\pm$ 0.0008 \\
\midrule
Metrics & \multicolumn{10}{c}{RMSE} \\
\cmidrule(lr){1-11}
Dataset & Emotion & Gambling & Language & Motor & Relational & Social & WM & Rest & Lorenz & HR \\
\midrule
MLP & 0.9500 \scriptsize $\pm$ 0.0011 & 0.9545 \scriptsize $\pm$ 0.0030 & 0.8928 \scriptsize $\pm$ 0.0029 & 0.9482 \scriptsize $\pm$ 0.0031 & 0.8499 \scriptsize $\pm$ 0.0098 & 0.9293 \scriptsize $\pm$ 0.0013 & 0.9224 \scriptsize $\pm$ 0.0096 & 0.9851 \scriptsize $\pm$ 0.0007 & 0.2444 \scriptsize $\pm$ 0.0052 & 0.7160 \scriptsize $\pm$ 0.0156 \\
Trans. & 0.3377 \scriptsize $\pm$ 0.0021 & 0.4051 \scriptsize $\pm$ 0.0031 & 0.3561 \scriptsize $\pm$ 0.0015 & 0.3218 \scriptsize $\pm$ 0.0019 & 1.0036 \scriptsize $\pm$ 0.0112 & 0.3553 \scriptsize $\pm$ 0.0024 & 0.4455 \scriptsize $\pm$ 0.0023 & 1.0676 \scriptsize $\pm$ 0.0023 & \underline{0.1492} \scriptsize $\pm$ 0.0040 & 0.7762 \scriptsize $\pm$ 0.0220 \\
LSTM & 0.3161 \scriptsize $\pm$ 0.0002 & 0.3740 \scriptsize $\pm$ 0.0004 & \underline{0.3494} \scriptsize $\pm$ 0.0005 & \underline{0.3116} \scriptsize $\pm$ 0.0004 & 1.2590 \scriptsize $\pm$ 0.0213 & 0.3247 \scriptsize $\pm$ 0.0001 & \underline{0.3658} \scriptsize $\pm$ 0.0018 & 1.2426 \scriptsize $\pm$ 0.0006 & 0.1525 \scriptsize $\pm$ 0.0030 & 0.5054 \scriptsize $\pm$ 0.0600 \\
STNDT & 0.8332 \scriptsize $\pm$ 0.0057 & 0.8882 \scriptsize $\pm$ 0.0043 & 0.9672 \scriptsize $\pm$ 0.0036 & 0.8701 \scriptsize $\pm$ 0.0025 & 0.9128 \scriptsize $\pm$ 0.0084 & 0.8999 \scriptsize $\pm$ 0.0010 & 0.8981 \scriptsize $\pm$ 0.0013 & 0.9653 \scriptsize $\pm$ 0.0007 & 0.5762 \scriptsize $\pm$ 0.0229 & 0.4945 \scriptsize $\pm$ 0.0665 \\
AMAG & \underline{0.3035} \scriptsize $\pm$ 0.0025 & \underline{0.2024} \scriptsize $\pm$ 0.0022 & 0.4140 \scriptsize $\pm$ 0.0015 & 0.5103 \scriptsize $\pm$ 0.0025 & \underline{0.5003} \scriptsize $\pm$ 0.0108 & \underline{0.2877} \scriptsize $\pm$ 0.0012 & 0.6180 \scriptsize $\pm$ 0.0032 & \underline{0.6104} \scriptsize $\pm$ 0.0012 & 0.8349 \scriptsize $\pm$ 0.0163 & \underline{0.2438} \scriptsize $\pm$ 0.0873 \\
LtrRNN & 0.8405 \scriptsize $\pm$ 0.0099 & 0.9497 \scriptsize $\pm$ 0.0043 & 1.0004 \scriptsize $\pm$ 0.0042 & 0.9325 \scriptsize $\pm$ 0.0042 & 0.9325 \scriptsize $\pm$ 0.0043 & 0.9663 \scriptsize $\pm$ 0.0009 & 0.9817 \scriptsize $\pm$ 0.0010 & 0.9966 \scriptsize $\pm$ 0.0011 & 1.0935 \scriptsize $\pm$ 0.0099 & 0.9721 \scriptsize $\pm$ 0.0901 \\
NetFormer & 0.7420 \scriptsize $\pm$ 0.0032 & 0.8548 \scriptsize $\pm$ 0.0037 & 0.9034 \scriptsize $\pm$ 0.0051 & 0.8169 \scriptsize $\pm$ 0.0018 & 0.8421 \scriptsize $\pm$ 0.0036 & 0.8209 \scriptsize $\pm$ 0.0008 & 0.8141 \scriptsize $\pm$ 0.0009 & 0.8913 \scriptsize $\pm$ 0.0005 & 0.5857 \scriptsize $\pm$ 0.0095 & 0.7124 \scriptsize $\pm$ 0.1034 \\
DCHO & \textbf{0.2172} \scriptsize $\pm$ 0.0017 & \textbf{0.1064} \scriptsize $\pm$ 0.0026 & \textbf{0.1891} \scriptsize $\pm$ 0.0007 & \textbf{0.1316} \scriptsize $\pm$ 0.0012 & \textbf{0.3093} \scriptsize $\pm$ 0.0094 & \textbf{0.1274} \scriptsize $\pm$ 0.0008 & \textbf{0.2491} \scriptsize $\pm$ 0.0016 & \textbf{0.3212} \scriptsize $\pm$ 0.0009 & \textbf{0.1254} \scriptsize $\pm$ 0.0010 & \textbf{0.0721} \scriptsize $\pm$ 0.0003 \\
\bottomrule
\end{tabular}
}
\caption{Prediction performance comparison (MAE and RMSE) of DCHO and baseline models on multiple datasets for raw fMRI signals forecasting with a 10-step horizon.}
\label{table2}
\end{table*}

\section{Results}
\subsection{Dataset and Preprocessing}
To evaluate the performance of our proposed method, we conducted experiments on both synthetic data and real-world fMRI datasets. The synthetic data were generated from two nonlinear dynamical systems: the Lorenz system \cite{wang2020adaptive} and the Hindmarsh–Rose (HR) neuronal model \cite{hindmarsh1984model}. In both cases, we simulated networks consisting of 20 coupled nodes.
For real-world evaluation, we utilized the Human Connectome Project (HCP) dataset \cite{van2012human}, which includes both resting-state and task-based fMRI recordings. Specifically, we employed seven task-based datasets: Emotion, Gambling, Language, Motor, Relational, Social, and Working Memory (WM) and a resting-state dataset (Rest). All HCP data had been preprocessed using the standard minimal preprocessing pipeline. We parcellated the cortex into 17 regions based on the Yeo 17-network atlas. It is important to note that the dimensionality of the higher-order brain connectivity tensor $H^t \in \mathbb{R}^{N \times N \times N}$ increases cubically with the number of brain regions N, indicating that the HOBC derived from the Yeo-17 parcellation is already highly complex and high-dimensional. For each subject and session, we extracted the average BOLD signal within each region and applied temporal standardization (z-scoring) independently for each region. 
In terms of data segmentation, all time series were divided into fixed-length temporal segment using a sliding window, with the window length set equal to the prediction length and a stride of 3. Following segmentation, 80\% of the data was used for training and the remaining 20\% for testing. 
See Appendix~\ref{Details of Datasets} and Appendix~\ref{Implementation Details} for further details.

\subsection{Evaluation of Higher-Order Brain Connectivity Tensor Prediction}
By predicting HOBC, DCHO encourages its latent representations to capture higher-order connectivity information. This facilitates both non-predictive and predictive downstream tasks. Before exploring these applications, we first evaluate DCHO’s performance in predicting HOBC.

\textbf{Experimental Setup: }Since HOBC prediction is a novel task, no existing methods have been specifically designed for this challenge. We therefore compare DCHO with commonly used baselines, including MLP \cite{rosenblatt1958perceptron}, LSTM \cite{graves2012long}, and Transformer \cite{vaswani2017attention}, to provide a meaningful reference. Detailed descriptions of these methods are provided in Appendix~\ref{Details of baseline}. We evaluate model performance using two common metrics (MAE and RMSE) across ten datasets (Emotion, Gambling, Language, Motor, Relational, Social, WM, Rest, Lorenz, HR).

\begin{figure}[tb]
\begin{center}
\centerline{\includegraphics[width=1.0\columnwidth]{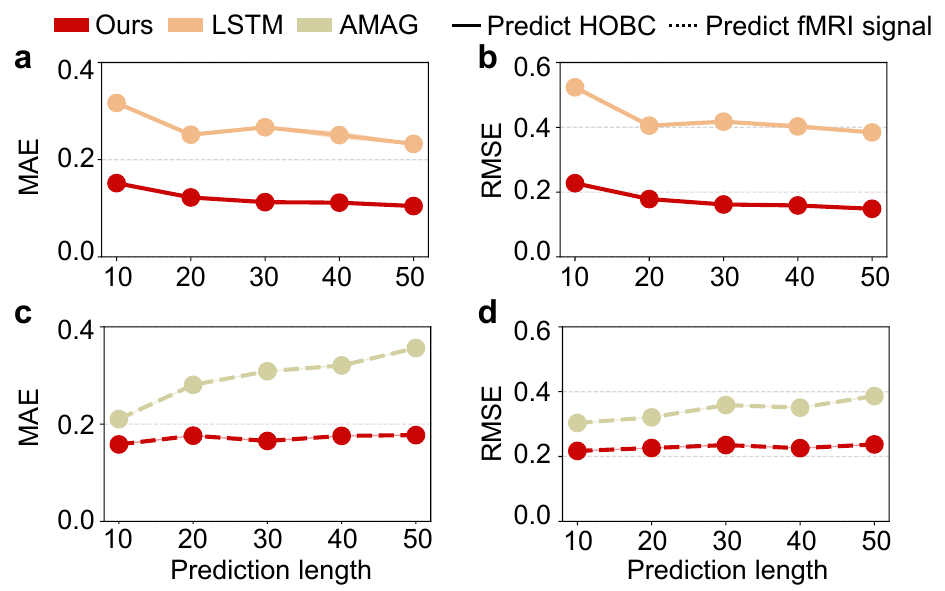}}
\caption{(a)-(b): Multi-step HOBC prediction performance of DCHO and LSTM on the Emotion dataset. (c)-(d): Multi-step raw fMRI signal prediction performance of DCHO and AMAG on the Emotion dataset. (Metrics: MAE (left) and RMSE (right),  Prediction lengths: 10 to 50 steps)}
\label{fig3}
\end{center}
\end{figure}

\textbf{Performance Analysis: }We systematically evaluate the model's performance from both short-term and long-term prediction perspectives. For short-term prediction, we conduct comparative experiments on ten datasets, with the baseline results summarized in Table \ref{table1}. The results show that DCHO consistently outperforms other baseline models in both MAE and RMSE, demonstrating superior modeling capability. For long-term prediction, as shown in Figure \ref{fig3} a–b, DCHO consistently maintains leading performance as the prediction horizon extends from 10 to 50 steps, demonstrating strong robustness. We attribute this to two key factors: (1) In the HOBC inference stage, DCHO leverages the dual-view encoder and the latent composition learner to effectively encode the latent information of HOBC; (2) In the latent trajectory prediction stage, DCHO introduces a latent-space prediction loss to model the dynamic evolution of HOBC in an abstract information space.

\begin{figure}[h]
\begin{center}
\centerline{\includegraphics[width=1.0\columnwidth]{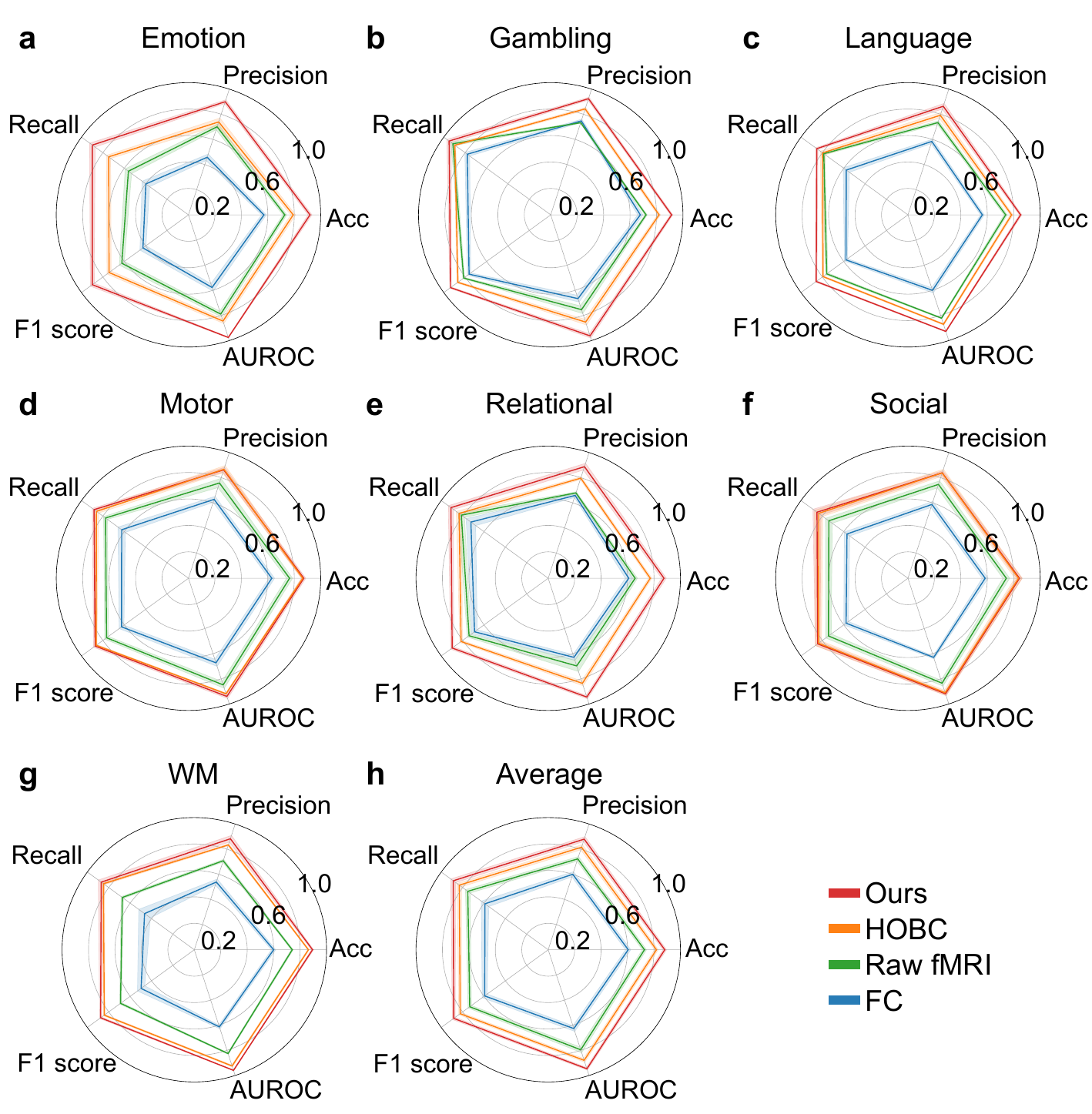}}
\caption{Comparison of task state classification performance across raw fMRI signals, FC, HOBC, and DCHO representations: DCHO outperforms raw fMRI signals, FC, and HOBC across seven cognitive task datasets and five metrics, as shown in (a–g), with average results in (h).}
\label{fig4}
\end{center}
\end{figure}

\subsection{Non-predictive Downstream Task}
Building upon its capacity to predict HOBC, DCHO learns rich information representations of HOBC. We used these higher-order representations to perform state classification across the seven HCP task datasets.

\textbf{Experimental Setup: }To validate the effectiveness of the representations extracted by the dual-view encoder, we compare them with three representative baselines: raw fMRI signals, FC, and HOBC. 
For each type of feature, we train the lightweight MLP classifiers separately, and evaluate performance using five standard metrics: Accuracy, Precision, Recall, F1 Score, and AUROC across seven datasets (Emotion, Gambling, Language, Motor, Relational, Social, WM).

\textbf{Performance Analysis: }As shown in Figure \ref{fig4}, our method achieves the best performance across all evaluation metrics on seven downstream tasks. In terms of average performance across all datasets, the representations extracted by DCHO outperform baseline methods, improving accuracy, precision, recall, F1 score, and AUROC by 6.24\%, 6.40\%, 5.58\%, 6.00\%, and 6.77\%, respectively. These results demonstrate the effectiveness of our learned representations.

\subsection{Predictive Downstream Task}
Compared to the study \cite{santoro2024higher, santoro2023higher} that focus primarily on analyzing HOBC within static or predefined temporal windows, DCHO captures the underlying dynamic evolution of HOBC, enabling more sophisticated predictive tasks.

\textbf{Benchmark Methods and Evaluation Metrics: }To evaluate the effectiveness of the pretrained dual-graph encoder and predictor, we compare DCHO against several state-of-the-art methods, including RNN-based models (LtrRNN \cite{pellegrino2023low}, LSTM \cite{graves2012long}), Transformer-based models (Transformer \cite{vaswani2017attention}, STNDT \cite{le2022stndt} and NetFormer \cite{lu2025netformer}), GNN-based models (AMAG \cite{li2023amag}), and a standard MLP \cite{rosenblatt1958perceptron} baseline. Detailed descriptions of these methods are provided in Appendix~\ref{Details of baseline}. We evaluate model performance using two common metrics (MAE and RMSE) across ten datasets (Emotion, Gambling, Language, Motor, Relational, Social, WM, Rest, Lorenz, HR). 

\textbf{Performance Analysis: }We systematically evaluate the model’s performance in raw fMRI signal prediction from both short-term and long-term prediction perspectives. For short-term prediction (with a forecast horizon of 10 steps), we conduct comparative experiments on ten datasets, with baseline results summarized in Table \ref{table2}. The results demonstrate that DCHO consistently outperforms existing models in terms of MAE and RMSE, showcasing its strong modeling capability. In the long-term prediction task, as the prediction horizon extends from 10 to 50 steps, DCHO maintains its leading performance, exhibiting notable robustness (Figure \ref{fig3} c-d). We attribute this advantage to two key factors: (1) Compared to baseline methods, the dual-graph encoder in DCHO effectively captures the latent information of HOBC, which inherently reflects nonlinear interactions across brain regions and enhances the modeling of brain dynamics; (2) In the latent space, the predictor captures the underlying evolutionary mechanisms of HOBC, enabling the modeling of deeper and more complex dynamic patterns, which facilitates more accurate long-range forecasting.

\subsection{Ablation Study}
We conduct an ablation study to evaluate the contributions of the proposed framework and its key components, with evaluations performed on both the task-based (Emotion) and resting-state datasets from HCP dataset. Specifically, we introduce the following model variants:
DCHO.a: An end-to-end baseline that directly predicts the future HOBC tensor $\hat{\mathbf{H}}^{t,t+T}$ from $\mathbf{X}^{t-T,t}$ and $\mathbf{A}^{t-T,t}$, without explicit structural decomposition.
DCHO.b: A variant that removes the local topological extractor from the dual-view encoder.
DCHO.c: A variant that removes the global topological extractor from the dual-view encoder.
DCHO.d: A variant that excludes the latent composition learner.
DCHO.e: A variant that replaces the latent-space prediction loss with the original HOBC prediction loss. As shown in Table \ref{ablation}, experimental results indicate that DCHO consistently outperforms all ablated variants, highlighting the effectiveness of the decomposition–composition framework, the dual-view encoder, and the latent composition learner in modeling HOBC dynamics.

\begin{table}[tb]
\centering
\small
\caption{Ablation Analysis}
{
\begin{tabular}{lcc}
\toprule
\textbf{Methods} & Emotion & Rest \\
\midrule
DCHO.a & 0.3556 $\pm$ 0.0009 & 0.4515 $\pm$ 0.0006  \\
DCHO.b & 0.3321 $\pm$ 0.0025 & 0.3317 $\pm$ 0.0004  \\
DCHO.c & 0.3770 $\pm$ 0.0013 & 0.3212 $\pm$ 0.0006  \\
DCHO.d & 0.4231 $\pm$ 0.0003 & 0.4825 $\pm$ 0.0005  \\
DCHO.e & 0.4432 $\pm$ 0.0011 & 0.4317 $\pm$ 0.0002  \\
DCHO   & \textbf{0.1744} $\pm$ 0.0004 & \textbf{0.1704} $\pm$ 0.0004 \\
\bottomrule
\end{tabular}
}
\label{ablation}
\end{table}

\section{Conclusion}
In this study, we propose DCHO, a novel framework for modeling and forecasting the temporal dynamics of higher-order brain connectivity (HOBC). By decomposing the complex prediction task into two manageable subtasks: HOBC tensor inference and latent trajectory prediction, DCHO effectively addresses the challenges of combinatorial explosion and high-dimensional modeling inherent to HOBC. The framework designs a dual-view encoder and a latent combinatorial learner to capture multiscale topological information and higher-order interaction patterns. Furthermore, it introduces a latent-space prediction loss to enable abstract and robust modeling of HOBC's temporal evolution. Extensive experiments on multiple neuroimaging datasets demonstrate that DCHO significantly outperforms existing state-of-the-art methods in both predictive tasks (e.g., brain dynamics forecasting) and non-predictive tasks (e.g., task-state classification), highlighting its strong performance and potential for neurocognitive modeling in real-world applications.

\textbf{Limitations and future work:} Although current DCHO architecture does not yet support cross-modal joint prediction, DCHO’s strong performance on fMRI and its modular decomposition–composition design provide a solid foundation for multimodal extension. Future work may explore joint modeling of fMRI and EEG to enhance performance in multimodal predictive tasks.

\section*{Acknowledgments}
This work was supported by the National Natural Science Foundation of China (62472206), National Key R\&D Program of China (2025YFC3410000), Shenzhen Science and Technology Innovation Committee (RCYX20231211090405003, KJZD20230923115221044), GuangDong Basic and Applied Basic Research Foundation (2025A1515011645 to ZC.L.), Shenzhen Doctoral Startup Project (RCBS20231211090748082 to XK.S.), Guangdong Provincial Key Laboratory of Advanced Biomaterials (2022B1212010003), and the open research fund of the Guangdong Provincial Key Laboratory of Mathematical and Neural Dynamical Systems, the Center for Computational Science and Engineering at Southern University of Science and Technology.

\bibliography{references}

\newpage

\appendix

\section{Algorithm}
\label{Algorithm}
We summarize the algorithm of DCHO as follows:
\begin{algorithm}
\caption{The Pipeline for HOBC Prediction}
\label{alg1} 
\textbf{Input:} The fMRI signals $\mathbf{X}^{t-T, t}$, FC matrices $\mathbf{A}^{t-T, t}$\\
\textbf{Output:} Predict HOBC $\hat{\mathbf{H}}^{t+1, t+T}$\\
\textbf{Training phase:}\\
\textbf{Step 1: HOBC inference}\\
(1) Encode $\mathbf{X}^{t-T, t}, \mathbf{A}^{t-T, t} \rightarrow \mathbf{Z}^{t-T, t}$ via the dual-view encoder\\
(2) Decode $\mathbf{Z}^{t-T, t} \rightarrow \hat{\mathbf{H}}^{t-T, t}$ via the higher-order decoder\\
(3) Train the encoder and decoder by minimizing inference loss: $loss_1 = ||\hat{\mathbf{H}}^{t-T, t} - \mathbf{H}^{t-T, t}||^2$\\
\textbf{Step 2: Latent trajectory prediction}\\
Freeze both the encoder and decoder, train the latent predictor to predict $\mathbf{Z}^{t-T, t} \rightarrow \hat{\mathbf{Z}}^{t+1, t+T}$ by minimizing the latent-space prediction loss: 
$loss_{2} = ||\hat{\mathbf{Z}}^{t+1, t+T} - \mathbf{Z}^{t+1, t+T}||^2.$\\
\textbf{Testing phase:}\\
(1) Encode input $\mathbf{X}^{t-T, t}, \mathbf{A}^{t-T, t} \rightarrow \mathbf{Z}^{t-T, t}$\\
(2) Predict: $\hat{\mathbf{Z}}^{t+1, t+T} = \text{Predictor}(\mathbf{Z}^{t-T, t})$\\
(3) Decode $\hat{\mathbf{Z}}^{t+1, t+T} \rightarrow \hat{\mathbf{H}}^{t+1, t+T}$ via decoder
\end{algorithm}

\section{The Proof of Theorem 1}
\label{The Proof of Theorem 1}
\paragraph{Proof.}
Add and subtract \( f_{\mathrm{dec}}(\mathbf{Z}^{t+1, t+T}) \), we have:
\begin{align}
& \bigl\|\hat{\mathbf{H}}^{t+1,t+T}-\mathbf{H}^{t+1,t+T}\bigr\|
   =\bigl\|\hat{\mathbf{H}}^{t+1,t+T} \nonumber\\
&-f_{\mathrm{dec}}(\mathbf{Z}^{t+1,t+T}) + f_{\mathrm{dec}}(\mathbf{Z}^{t+1,t+T}) - \mathbf{H}^{t+1,t+T}
     \bigr\|.
\end{align}
By the triangle inequality, it yields:
\begin{align}
& \bigl\|\hat{\mathbf{H}}^{t+1,t+T}-\mathbf{H}^{t+1,t+T}\bigr\|
   \le \bigl\|\hat{\mathbf{H}}^{t+1,t+T} \nonumber\\
&-f_{\mathrm{dec}}(\mathbf{Z}^{t+1,t+T})\bigr\| + \bigr\|f_{\mathrm{dec}}(\mathbf{Z}^{t+1,t+T}) - \mathbf{H}^{t+1,t+T}
     \bigr\|.
\end{align}
According to \( f_{\mathrm{dec}} \) is \( L_{\mathrm{dec}} \)-Lipschitz, it follows that:
\begin{align}
&\bigl\|\hat{\mathbf{H}}^{t+1,t+T} - f_{\mathrm{dec}}(\mathbf{Z}^{t+1,t+T})\bigr\|
\;\le\;
L_{\mathrm{dec}}\,
\bigl\|\hat{\mathbf{Z}}^{t+1,t+T}\nonumber\\
&-\mathbf{Z}^{t+1,t+T}\bigr\|
=
L_{\mathrm{dec}}\,
\epsilon_{\mathrm{dyn}}.
\end{align}
Combining the inference error: 
\begin{equation}
\epsilon_{\mathrm{inf}}=\bigl\|f_{\mathrm{dec}}\bigl(f_{\mathrm{enc}}(\mathbf{X}^{t+1, t+T}, \mathbf{A}^{t+1, t+T})\bigr) - \mathbf{H}^{t+1, t+T}\bigr\|.
\end{equation}
we can obtain the following:
\begin{equation}
\bigl\|
    \hat{\mathbf{H}}^{t+1, t+T}
    -
    \mathbf{H}^{t+1, t+T}
\bigr\| \le \epsilon_{\mathrm{inf}} + L_{\mathrm{dec}}\cdot \epsilon_{\mathrm{dyn}}.
\end{equation}

\section{Details of Datasets}
\label{Details of Datasets}

\subsection{Synthetic Data}

The \textbf{Lorenz system} is a three-variable chaotic dynamical system originally derived from a simplified model of atmospheric convection. It is given by:

\begin{equation*}
\begin{aligned}
\frac{dx}{dt} &= \sigma (y - x), \\
\frac{dy}{dt} &= x(\rho - z) - y, \\
\frac{dz}{dt} &= xy - \beta z,
\end{aligned}
\end{equation*}

where $\sigma$, $\rho$, and $\beta$ are system parameters. For the classic chaotic regime ($\sigma$ = 10, $\rho$ = 28, $\beta$ = 8/3), the system exhibits sensitive dependence on initial conditions and generates the well-known Lorenz attractor.

The \textbf{Hindmarsh–Rose (HR) system} is a three-dimensional system of nonlinear ordinary differential equations used to simulate neuronal spiking and bursting activity. It is defined as:

\begin{equation*}
\begin{aligned}
\frac{dx}{dt} &= y - a x^3 + b x^2 - z + I, \\
\frac{dy}{dt} &= c - d x^2 - y, \\
\frac{dz}{dt} &= r \left[ s(x - x_R) - z \right],
\end{aligned}
\end{equation*}

where $x$ represents the membrane potential, $y$ the fast recovery variable, and $z$ the slow adaptation current. Parameters $a$, $b$, $c$, $d$, $r$, $s$, $x_R$, and the external current $I$ control the dynamic regimes of the neuron, including periodic spiking and chaotic bursting. A commonly used parameter setting is $a = 1$, $b = 3$, $c = 1$, $d = 5$, $r = 0.006$, $s = 4$, $x_R = -1.6$, with $I$ as the external input current.

\subsection{HCP dataset}

The \textbf{Human Connectome Project (HCP)} is a large-scale neuroimaging initiative that provides high-resolution structural and functional MRI data from healthy adult participants. The dataset includes both resting-state fMRI and task-based fMRI recordings, enabling the study of brain connectivity and cognitive function under various conditions.
The task-based fMRI data cover \textbf{seven cognitive domains}, each designed to elicit specific neural responses:

\begin{itemize}
    \item \textbf{Emotion}: (emotional face matching)
    \item \textbf{Gambling}: (reward processing)
    \item \textbf{Language}: (story comprehension and math)
    \item \textbf{Motor}: (movement execution)
    \item \textbf{Relational}: (relational reasoning)
    \item \textbf{Social}: (theory of mind)
    \item \textbf{Working Memory}: (n-back task)
\end{itemize}

In addition to these tasks, the HCP dataset includes \textbf{resting-state fMRI} sessions, where participants are scanned while not performing any explicit task.

\section{Details of baseline}
\label{Details of baseline}

The details of baselines are elaborated as follows:
\begin{itemize}
    \item \textbf{MLP}~\cite{rosenblatt1958perceptron}: A standard multi-layer perceptron that models temporal dynamics, without considering spatial or temporal structure explicitly.
    
    \item \textbf{LSTM}~\cite{graves2012long}: A recurrent neural network that captures temporal dependencies via gating mechanisms, widely used in sequence modeling tasks.
    
    \item \textbf{Transformer}~\cite{vaswani2017attention}: A self-attention-based model that captures long-range dependencies in time series, without relying on recurrence.

    \item \textbf{STNDT}~\cite{le2022stndt}: A Transformer-based method, STNDT captures spatiotemporal neural dynamics and enhances prediction through contrastive learning.
    
    \item \textbf{AMAG}~\cite{li2023amag}: A graph-based method, AMAG uses additive and multiplicative message passing to adaptively model pairwise interactions and forecast neural activity.

    \item \textbf{LtrRNN}~\cite{pellegrino2023low}: A tensor-based model that uncovers low-dimensional neural dynamics by fitting RNNs to large-scale cortical recordings during learning tasks.
    
    \item \textbf{NetFormer}~\cite{lu2025netformer}: A transformer-based model designed for interpretable prediction of nonstationary neuronal dynamics and time-varying connectivity patterns from population activity.

\end{itemize}

\section{Implementation Details}
\label{Implementation Details}

We implemented all baseline methods and the proposed DCHO model using PyTorch. Our dual-graph encoder consists of two parallel branches designed to extract local and global topological features, respectively. Each branch consists of a two-layer Graph Network. The latent representation dimension is set to 32. The dynamics predictor is implemented as a four-layer Long Short-Term Memory (LSTM) network, where all layers share the same input and hidden dimensions. The dual-stream higher-order decoder comprises spatial and temporal branches, each constructed using a two-layer TransformerEncoder. Each TransformerEncoderLayer utilizes 4 attention heads, a feedforward dimension twice the latent size, and GELU activation.

Model training was performed using the Adam optimizer with an initial learning rate of $1\times10^{-3}$, weight decay of $1\times10^{-6}$, and a dropout rate of 0.1. Training proceeded for 200 epochs with a batch size of 8. Each experimental result was repeated 10 times, and the mean and standard deviation were calculated.

We prepared data from both real-world and synthetic sources for training and evaluation. For the seven task-based and one resting-state dataset from the Human Connectome Project (HCP), we selected 10 subjects per dataset. Each subject's fMRI time series was segmented into fixed-length temporal windows, with 80\% of the data used for training and 20\% for testing. For synthetic data, we generated 10 trajectories from the Lorenz system and 10 from the Hindmarsh–Rose (HR) model, each with 500 time steps, and partitioned them similarly into 80\% training and 20\% testing splits.

All experiments were conducted on a high-performance workstation equipped with an NVIDIA RTX 4090 GPU (24GB VRAM), AMD EPYC 7T83 CPU, and 512GB RAM, running Ubuntu 22.04 with Python 3.9.20 and CUDA 12.4. This computational environment ensured training efficiency and reproducibility across all experimental runs.

\section{Additional ablation experiments}
To further validate the effectiveness of our method, we include additional ablation experiments. DCHO.f is a variant that replaces the dual-view encoder with an MLP. DCHO.g substitutes the LSTM with a Transformer, and DCHO.h replaces the LSTM with a GRU.  As shown in Table \ref{Additional ablation}, the inferior performance of DCHO.f further confirms the effectiveness of the dual-view encoder. In addition, both DCHO.g and DCHO.h exhibit only slight performance degradation, which further verifies that our model’s gains mainly derive from the decomposition–composition strategy (i.e., the latent prediction loss) rather than from the choice of predictor architecture.

\begin{table}[ht]
\centering
\small
\caption{Additional ablation analysis}
{
\begin{tabular}{lcc}
\toprule
\textbf{Methods} & Emotion & Rest \\
\midrule
DCHO.f & 0.3849 $\pm$ 0.0019 & 0.3527 $\pm$ 0.0013  \\
DCHO.g & 0.1841 $\pm$ 0.0007 & 0.1725 $\pm$ 0.0011  \\
DCHO.h & 0.1962 $\pm$ 0.0010 & 0.1812 $\pm$ 0.0008  \\
DCHO   & \textbf{0.1744} $\pm$ 0.0004 & \textbf{0.1704} $\pm$ 0.0004 \\
\bottomrule
\end{tabular}
}
\label{Additional ablation}
\end{table}

\section{Expanded Experimental Evaluation with Additional Datasets and Baselines}
To further evaluate the performance of our model, we additionally include more baseline methods (HGFM, ALTER, and NEUROTREE) as well as a new dataset (the resting-state subset of CHCP). These baselines were designed based on graph architectures but focus primarily on diagnostic tasks rather than dynamic HOBC prediction, and therefore cannot be directly compared under our task setting. To enable a fair and appropriate comparison, we adapt HGFM, ALTER, and NEUROTREE to our experimental environment. As shown in Table \ref{Additional experimental evaluation}, all three baselines exhibit lower performance than our model.

\begin{table}[ht]
\centering
\small
\caption{Additional experimental evaluation}
{
\begin{tabular}{lccc}
\toprule
\textbf{Methods} & Emotion (HCP) & Rest (CHCP) \\
\midrule
HGFM & 0.3721 $\pm$ 0.0007 & 0.2324 $\pm$ 0.0016 \\
ALTER & 0.3905 $\pm$ 0.0021 & 0.2652 $\pm$ 0.0017 \\
NEUROTREE & 0.4169 $\pm$ 0.0014 & 0.2360 $\pm$ 0.0014 \\
DCHO   & \textbf{0.1744} $\pm$ 0.0004 & \textbf{0.2114} $\pm$ 0.0010 \\
\bottomrule
\end{tabular}
}
\label{Additional experimental evaluation}
\end{table}

\section{Complexity analyses}

Our method consists of two stages: the first stage infers HOBC, and the second stage performs latent trajectory prediction. Since the second stage operates in a low-dimensional latent space, its computational cost is negligible. Therefore, we focus on the complexity of the first stage. The main cost in the first stage comes from the dual-view encoder and the high-order decoder. The dual-view encoder has two parallel branches: a local topology extractor and a global topology extractor.

Local branch complexity comprises four parts: adjacency normalization: $O(BTN^{2})$, linear projection: $O(BTNF^{2})$, cosine similarity: $O(BTN^{2})$, neighbor aggregation: $O(BTN^{2}F)$. Summing yields
\begin{equation}
  O\!\left(BT\,(N^{2}F + NF^{2} + N^{2})\right)
  \approx O(BTNF^{2}),
\end{equation}
since $NF^{2} > N^{2}F > N^{2}$ in our setting. The global branch includes the normalized Laplacian construction $O(BTN^{2})$, the Chebyshev recurrence $O(BTN^{2}F)$, and the Chebyshev linear mapping $O(BTNF^{2})$. Summing these terms gives
\begin{equation}
  O\!\left(BT\,(N^{2} + N^{2}F + NF^{2})\right)
  \approx O(BTNF^{2}).
\end{equation}
Since the two branches run in parallel, the total complexity is determined by the dominant term among them. Therefore, the overall complexity of the dual-view encoder is $O(BTNF^{2})$.

The high-order decoder consists of five components: linear projection: $O(BTNF^{2})$, latent composition learner: $O(BTMF^{2})$, spatial Transformer: $O(BTM^{2}F)$, temporal Transformer: $O(BTMF^{2})$, output projection: $O(BTMF^{2})$. Summing these terms gives
\begin{equation}
  O\!\left(BT\,(NF^{2} + MF^{2} + M^{2}F)\right)
  \approx O(BTM^{2}F),
\end{equation}
since $M^{2}F > MF^{2} > NF^{2}$.

Therefore, the overall complexity of the dual-view encoder combined with the high-order decoder amounts to $O(BTM^{2}F)$. Notably, for the downstream tasks, we only employ the dual-view encoder, reducing the practical complexity to $O(BTNF^{2})$.

\section{Interpretability analyses}
Our method achieves the largest gains in Emotion, Gambling, Language, and Working Memory, indicating that the learned embedding is particularly effective at capturing higher-order cognitive processes. These tasks rely on distributed cross-network interactions, latent affective or value states, semantic integration, and sustained executive control. The consistent improvements suggest that our embedding encodes a more abstract and cognitively aligned representation—one that goes beyond local activation or static connectivity and instead reflects the dynamic, integrative computations underlying advanced human cognition.

\end{document}